\documentclass[prb,amsmath,amssymb,lengthcheck,showpacs,floatfix,groupedaddress,superscriptaddress,longbibliography]{revtex4-1}
\usepackage{graphicx}
\usepackage[english]{babel}
\usepackage{times}
\usepackage{dcolumn}
\usepackage[usenames,dvipsnames]{color}
\usepackage{units}
\usepackage{bm}
\usepackage{soul}
\usepackage[utf8]{inputenc}

\begin{document}

\title{Reversal of particle-hole scattering-rate asymmetry in 
Anderson impurity model}

\author{R. \v{Z}itko}

\affiliation{Jo\v{z}ef Stefan Institute, Jamova 39, SI-1000 Ljubljana, Slovenia}
\affiliation{Faculty  of Mathematics and Physics, University of Ljubljana,
Jadranska 19, SI-1000 Ljubljana, Slovenia}

\author{H. R. Krishnamurthy}
\affiliation{Department of Physics, Center for Condensed Matter Theory,
Indian Institute of Science, Bengaluru - 560012, India}

\author{B. Sriram Shastry}
\affiliation{Physics Department, University of California, Santa Cruz,
CA 95064}

\date{\today}

\begin{abstract}
We study the particle-hole asymmetry of the scattering rate in
strongly correlated electron systems by examining the cubic $\omega^3$
and $\omega T^2$ terms in the imaginary part of the self-energy of the
Anderson impurity model. We show that the sign is opposite in the
weak-coupling and strong-coupling limits, explaining the differences
found in theoretical approaches taking the respective limits as the
starting points. The sign change in fact precisely delineates the
cross-over between the weak and strong correlation regimes of the
model. For weak interaction $U$ the sign reversal occurs for small
values of the doping $\delta=1-n$, while for interaction of order $U
\approx 2 \Gamma$, $\Gamma$ being the hybridization strength, the
cross-over curve rapidly shifts to the large-doping range. This curve
based on the impurity dynamics is genuinely different from other
cross-over curves defined through impurity thermodynamic and
static properties.
\end{abstract}


\maketitle

\newcommand{\vc}[1]{{\mathbf{#1}}}
\renewcommand{\Im}{\mathrm{Im}}
\renewcommand{\Re}{\mathrm{Re}}

\newcommand{\expv}[1]{\langle #1 \rangle}
\newcommand{\ket}[1]{| #1 \rangle}
\newcommand{\Tr}{\mathrm{Tr}}

In contemporary strongly correlated quantum materials, such as the
cuprate superconductors and sodium cobaltates, one finds that spectral
line shapes from angle-resolved photoemission spectroscopy (ARPES)
differ qualitatively from those in simple Fermi liquids. The origin of
the difference has been traced to a large correlation-induced
asymmetry in the imaginary part of the self-energy
\cite{gweon2011,shastry2012}, which can be expanded as
\begin{equation}
\Im \Sigma(\omega,T) = a(\omega^2 + \pi^2 T^2) + b \omega^3 + c \omega
T^2 + \ldots
\end{equation}
For context note that the usually quoted Fermi-liquid self-energy,
namely the first two terms in this expression, is {\em even} in
$\omega$. While this is dominant at the lowest energies, the higher
order {\em odd} in $\omega$ terms become important when their
coefficients $(b,c)$ become sufficiently large. This is found to
happen in the strong-correlation models, while in weakly correlated
systems these coefficients are very small. The signs of $(b,c)$ are of
particular importance: they determine whether particles or holes have
the shorter lifetime. Since $a<0$, if $b<0$ the particle-like excitations
scatter more strongly on the impurity (are more damped) than the
hole-like excitations with the same excitation energy (absolute value
of $\omega$), and vice-versa for $b>0$. Understanding the asymmetry of
the self-energy is a problem of great current interest. The asymmetry
of $\Im \, \Sigma$ is relevant to transport coefficients such as
the thermopower, where the entropy and charge are carried by both
particle-like excitations above the Fermi level and hole-like
excitations below it. However in the low-$T$ thermopower there are
other competing factors (asymmetry of the density of states, asymmetry
of the quasiparticle velocities), hence the situation is not solely
controlled by the sign of the scattering asymmetry.

\begin{figure}[htbp]
\centering
\includegraphics[width=0.8\columnwidth]{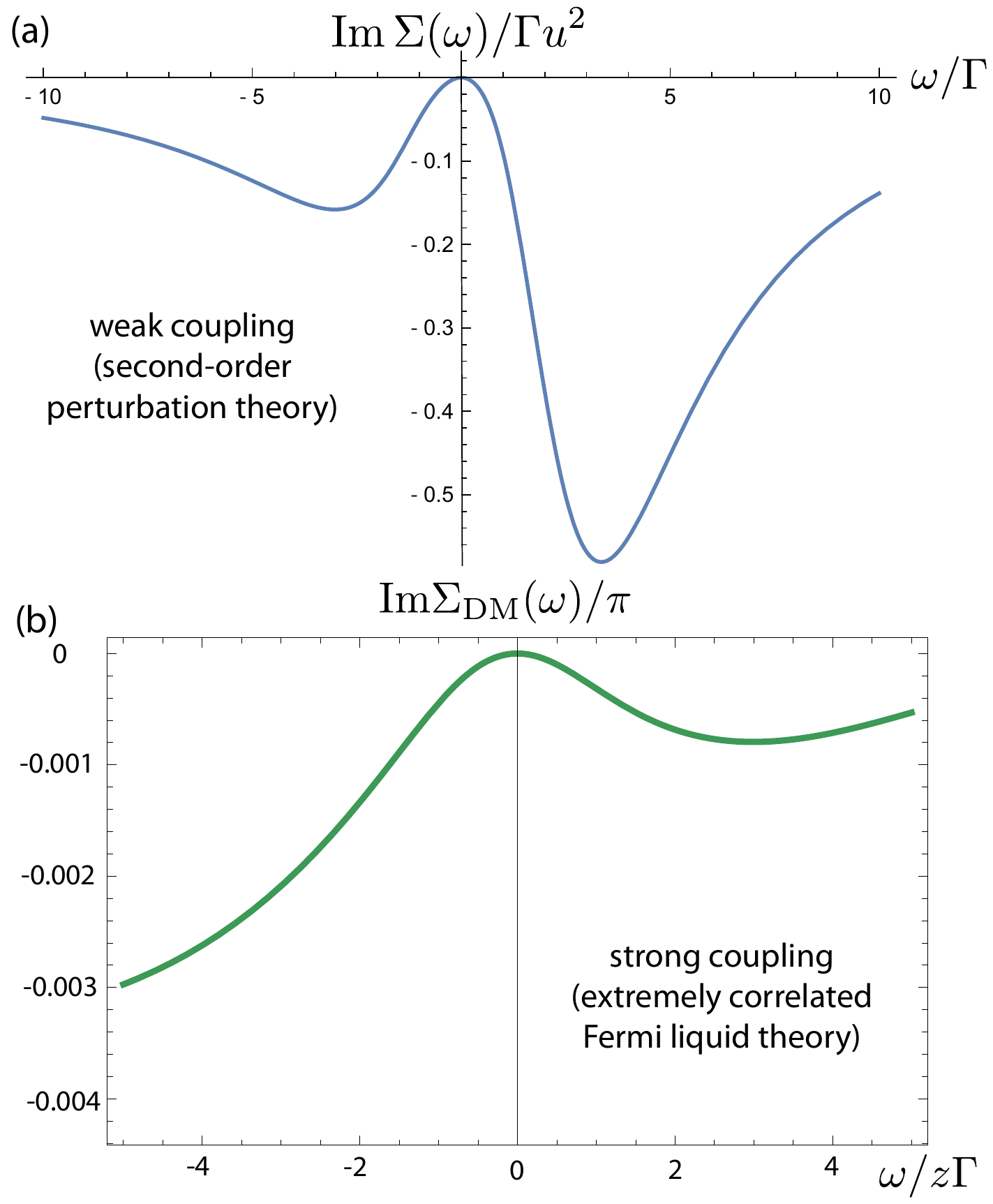}
\caption{Imaginary parts of the self-energy in (a) weak-coupling
theory and in (b) strong-coupling theory have the opposite sign of
the particle-hole asymmetry. The second-order perturbation theory
result corresponds to $E_d/\Gamma=1$ in Fig.~2 of
Ref.~\onlinecite{horvatic1980}, the extremely-correlated Fermi liquid
theory result to $n=0.6$ in Fig.~5 of Ref.~\onlinecite{shastry2013}.}
\label{intro}
\end{figure}

The single impurity Anderson model (SIAM) is a ``laboratory example''
of an exactly solvable many-body problem. It is simpler but has many
similarities to the lattice many-body problems such as the Hubbard
model. It is therefore a natural place to understand the magnitude and
signs of the asymmetric corrections to the lowest-order Fermi liquid
theory result mentioned above. The goal of this paper is to explore
this asymmetry by using the numerical renormalization group (NRG), and
to contrast it with various approximate theories. We report a
surprising result in this well-studied problem: we find a line in the
$U$-$n$ plane where the asymmetry changes sign. Here $U$ is the
interaction strength and $n$ the impurity occupancy. Along this
1-dimensional line in $U$-$n$ plane, the particle-hole symmetry of the
scattering rate is exactly fulfilled up to the fifth and higher order
terms. This change of sign demarcates the border between the
qualitatively different regimes of weak and strong correlations.
Indeed, this work was motivated by the puzzling observation that
weak-coupling approaches (e.g., perturbation theory in the interaction
strength $U$) and strong-coupling techniques (e.g., the extremely
correlated Fermi liquid (ECFL) theory
\cite{shastry2011,shastry2012,shastry2013}) give opposite signs for
the asymmetry, as illustrated in Fig.~\ref{intro}.

The SIAM is defined by the Hamiltonian
\begin{equation}
H=\epsilon_d n + U n_{\uparrow} n_{\downarrow}
+ \sum_{k\sigma} \epsilon_k c^\dag_{k\sigma} c_{k\sigma}
+ \sum_{k\sigma} \left( t_k c^\dag_{k\sigma} d_\sigma + \text{h.c.}
\right)
\end{equation}
where $\epsilon_d$ is the impurity level, $d_\sigma$ are impurity
operators, $n_\sigma=d^\dag_{\sigma} d_\sigma$ and
$n=n_\uparrow+n_\downarrow$, $c_{k\sigma}$ are operators for
conduction-band electrons with energy $\epsilon_k$ that couple with
the impurity with amplitude $t_k$. The hybridisation strength is
$\Gamma=\pi \sum_k |t_k|^2 \delta(\omega-\epsilon_k)$; we will assume
it to be a constant function in the domain $-D<\omega<D$. In the
following we will make use of the Hartree-Fock parameter
$E_d=\epsilon_d+U \expv{n_\sigma}$ with $E_d=0$ corresponding to the
p-h symmetric $n=1$ case. We also define the dimensionless interaction
$u=U/\pi\Gamma$. We will limit our consideration to $n<1$, since the
results for $n>1$ can be obtained by the p-h transformation $d_\sigma
\to d^\dag_\sigma$, $c_{k\sigma} \to -c_{k\sigma}^\dag$ which takes
$\omega$ to $-\omega$.

Second-order perturbation theory in $U$ for the SIAM
\cite{horvatic1980} (e.g. Fig.~\ref{intro}(a) for $E_d/\Gamma=1$)
predicts $b$ to be of constant sign as a function of $n$ in the full
domain $0<n<1$, with zero value at $n=1$. Specifically, $b<0$, i.e.
the particle-like excitations scatter more strongly. At non-zero but
low temperatures, the asymmetry of $\Im\Sigma(\omega)$ at low $\omega$
will be controlled by the $\omega T^2$ term. Due to conformal symmetry
of the FL fixed point, $c = b \pi^2$, thus the sign of the asymmetry
does not depend on T at low enough temperatures \cite{horvatic1982}.

\begin{figure}[htbp]
\centering
\includegraphics[width=\columnwidth]{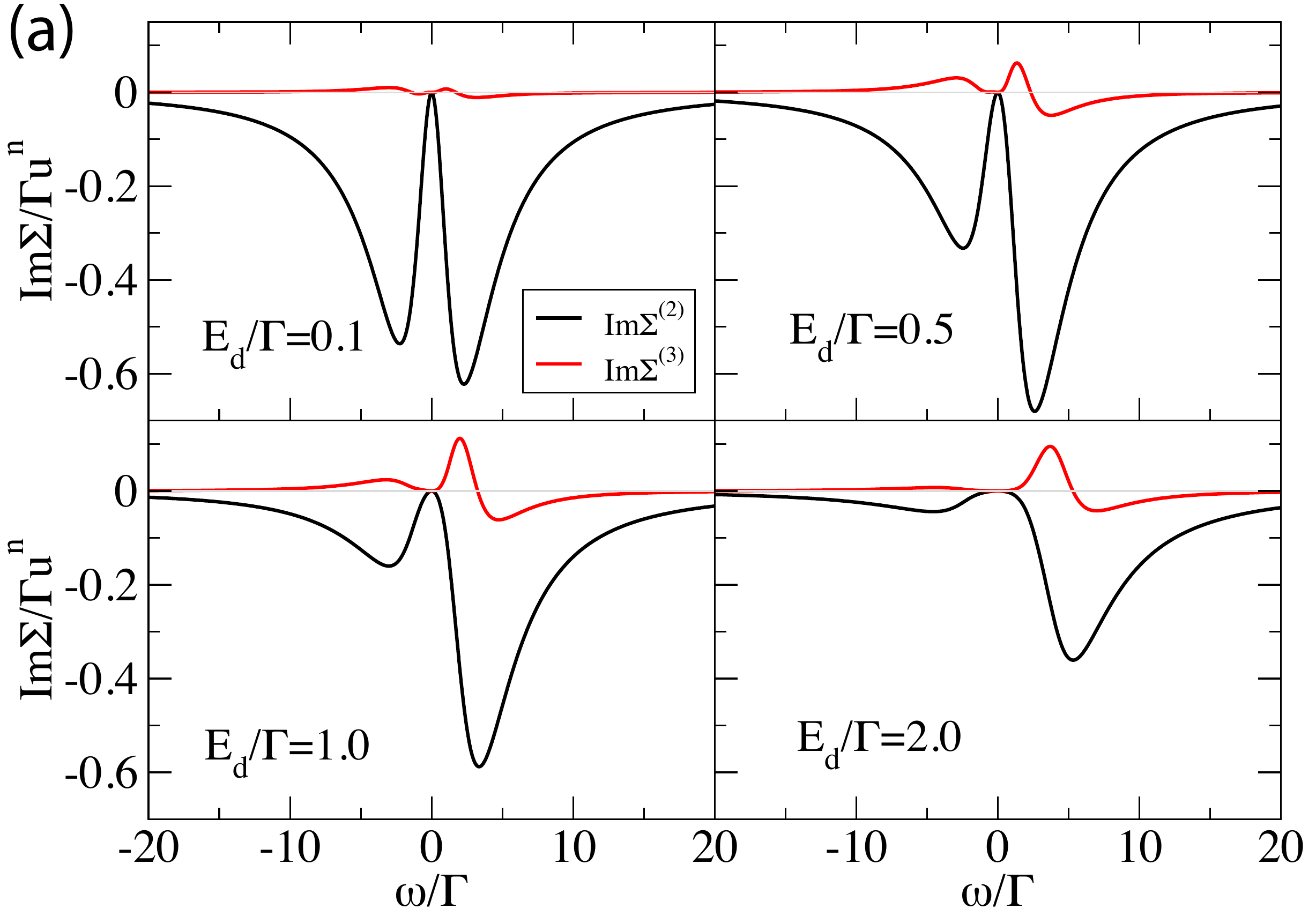}
\includegraphics[width=\columnwidth]{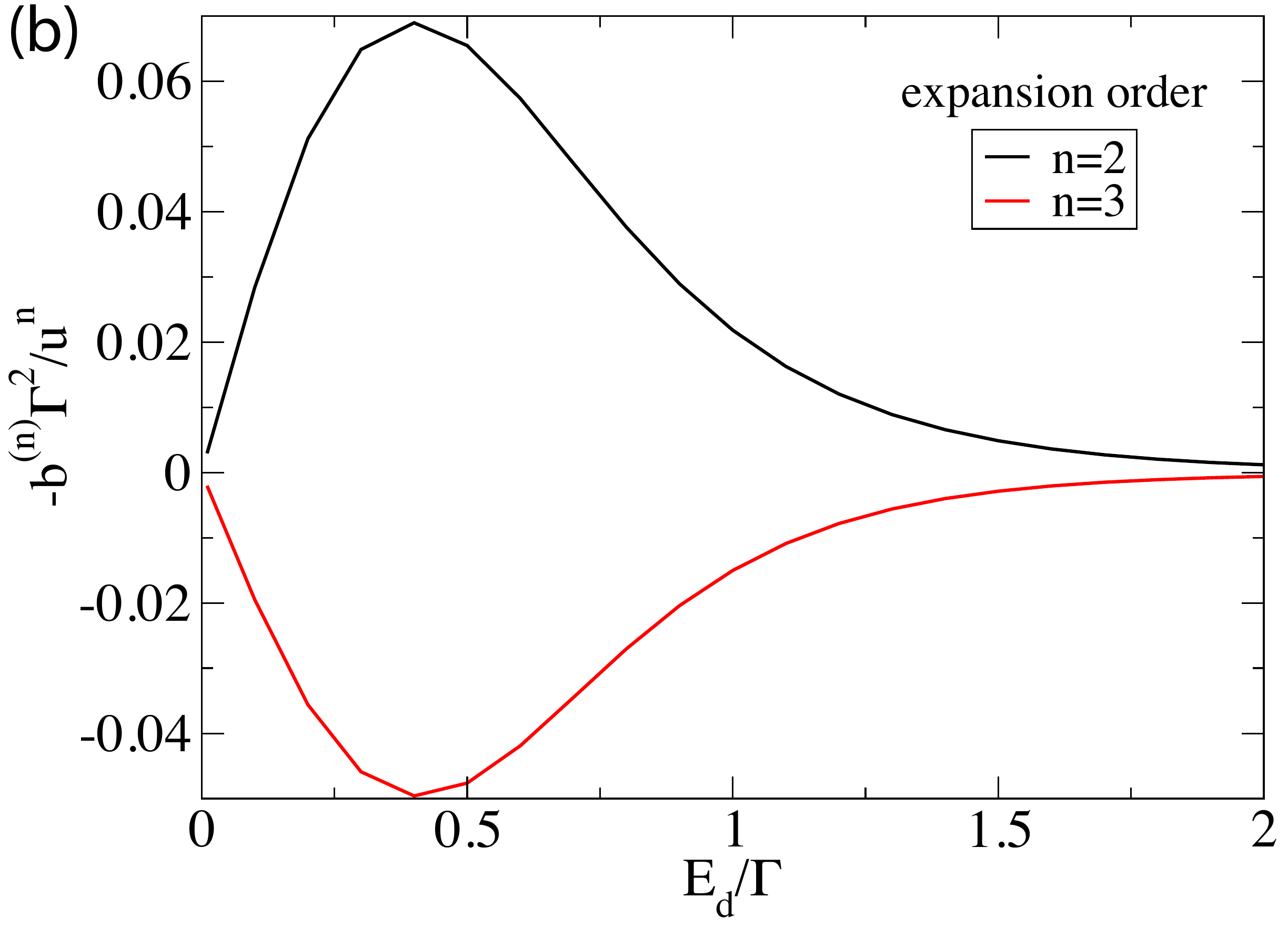}
\caption{(a) Second-order and third-order imaginary part of the
self-energy for a range of the Hartree-Fock parameters
$E_d=\epsilon_d+U\expv{n_\sigma}$. The y-axis is scaled as $1/(\Gamma
u^n)$ where $n$ is the expansion order and $u=U/\pi\Gamma$. $T=0$.
(b) Coefficient $b$ of the $\omega^3$ term in
$\Im\,\Sigma(\omega,T=0)$ at second and third order
in perturbation theory.} \label{comp}
\end{figure}

By extending the perturbative expansion to third order, we find that
the third-order contributions tend to have the opposite sign of the
second-order ones for small $\omega$, i.e., they reduce the
scattering. Furthermore, the asymmetry of the third-order contribution
is such that the $\omega>0$ part is dominant (for $n<1$), thus the
third-order reduction in scattering is stronger for particle-like
excitations, see Fig.~\ref{comp}(a). There is thus a competition
between the second and third-order contributions which may lead to a
change in sign of the asymmetric terms. The contributions $b^{(2)}$
and $b^{(3)}$ to the coefficient $b$ actually follow very similar
qualitative $E_d$ dependence, see Fig.~\ref{comp}(b), except for the
sign and evidently a different power of $u=U/\pi\Gamma$. Based on
these results, the change of sign in $b$ should occur for $u\sim1$,
i.e., $U \sim \pi\Gamma$. Since the perturbation theory also breaks
down at $u\sim1$, we cannot make any more precise statement about the
details of this sign change within the bare perturbative approach.

\begin{figure}[htbp]
\centering
\includegraphics[width=\columnwidth]{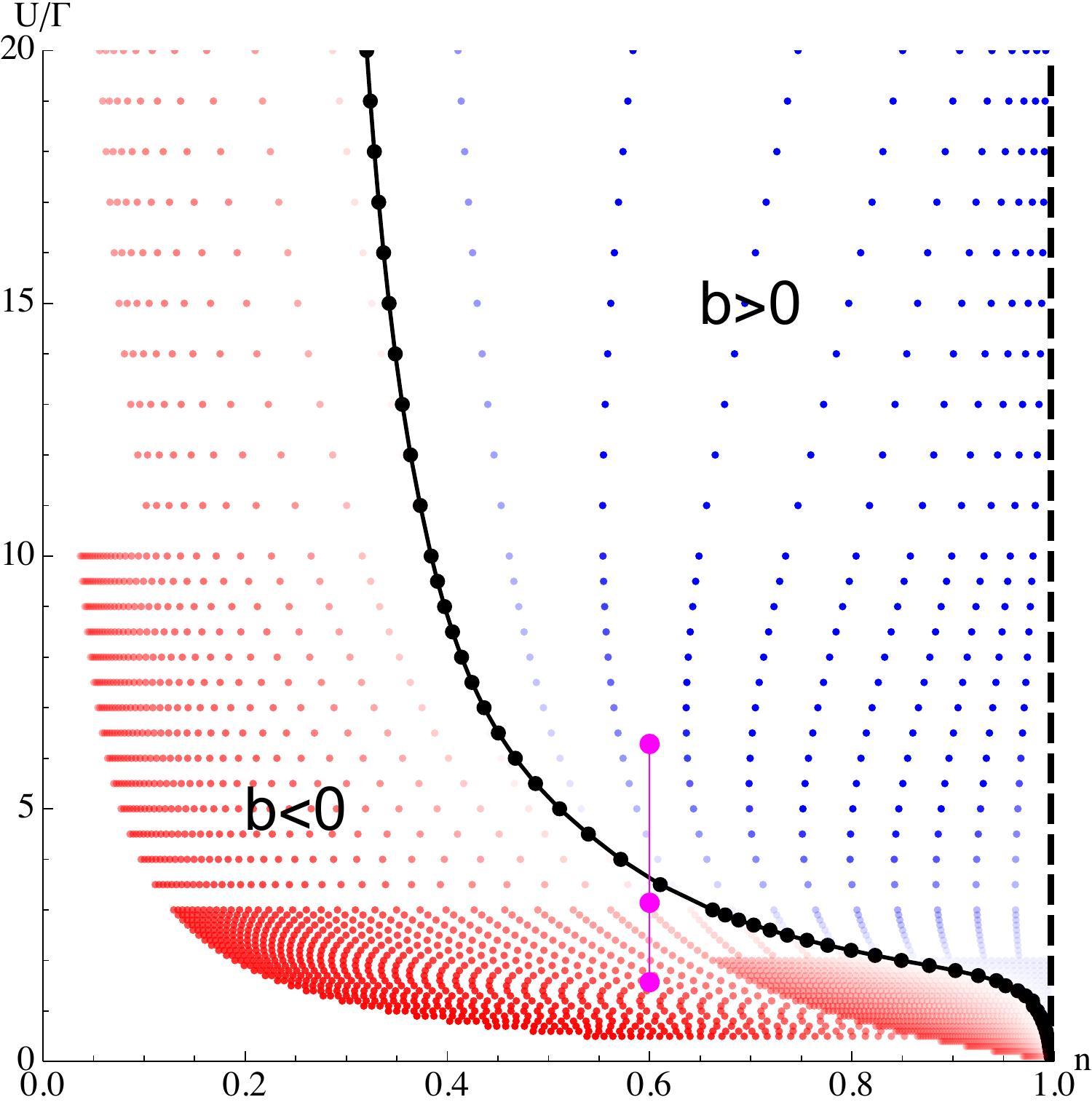}
\includegraphics[width=0.49\columnwidth]{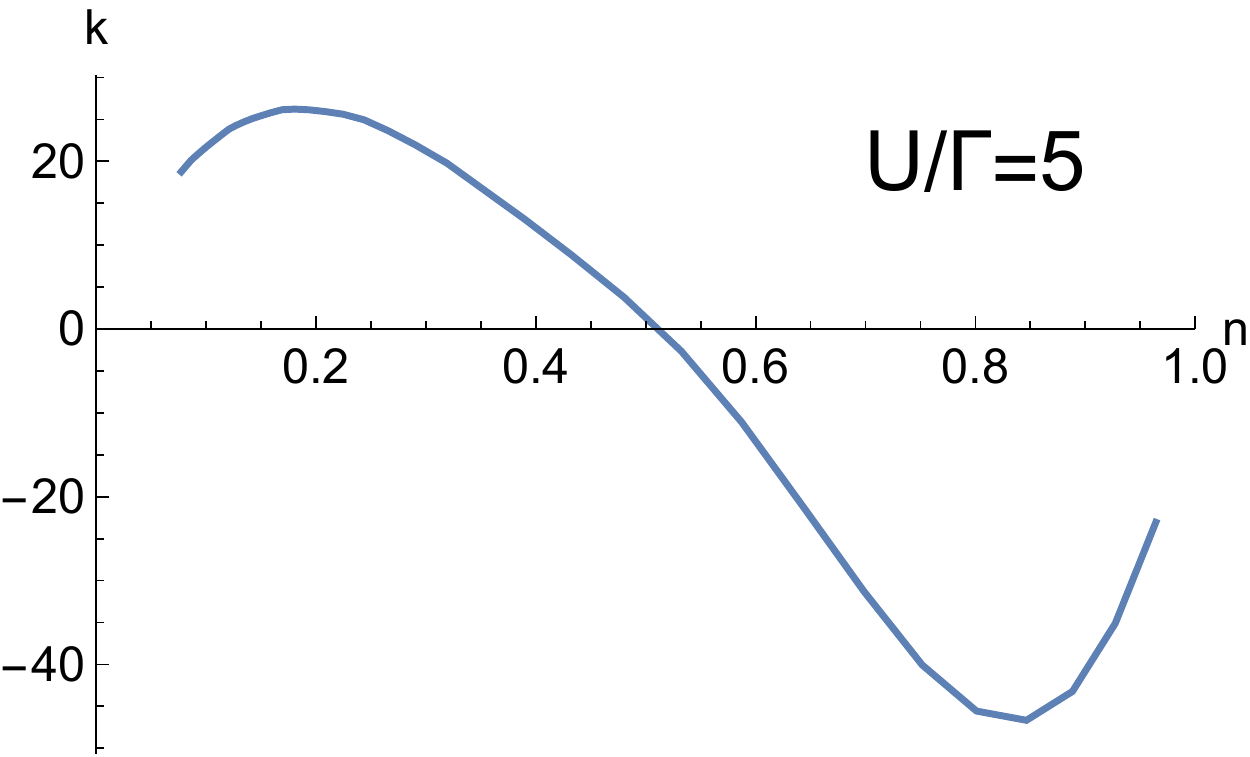}
\includegraphics[width=0.49\columnwidth]{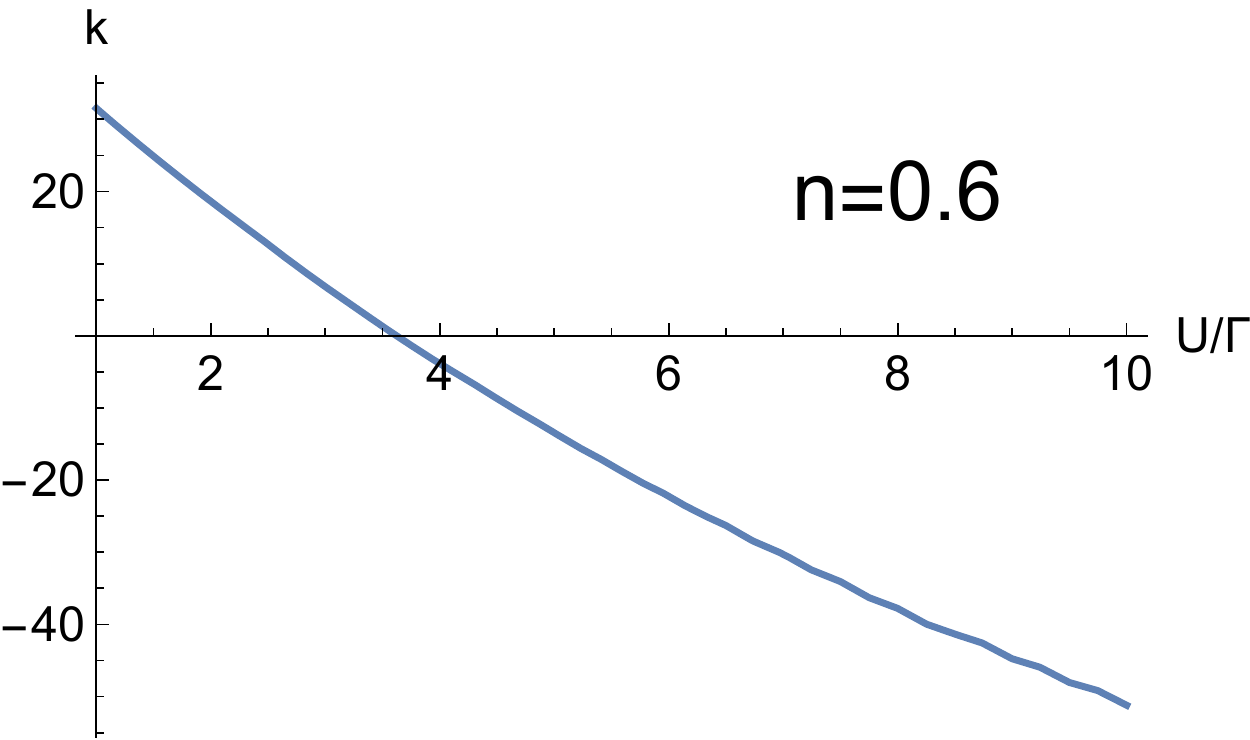}
\caption{Top: Sign (red negative, blue positive) and magnitude (color
saturation) of the $\omega^3$ term in $\Im\Sigma(\omega,T=0)$. The
quantity shown is the coefficient $k$ in the fit of the
antisymmetrized and normalized combination
$[\Im\Sigma(\omega)-\Im\Sigma(-\omega)]/[\Im\Sigma(\omega)+\Im
\Sigma(-\omega)]$ with the linear function $k \omega$. The fit is
performed in an energy interval $\omega \in [-\xi:\xi]$; here $\xi$ is
the low-energy scale of the problem defined as the temperature where
the impurity moment is screened (and is equivalent to the Kondo
temperature in the Kondo regime of the model). Notice that $k \approx
b/a$ and that $a<0$. The three magenta points joined by a line are
considered in Fig.~\ref{diff}. The dashed lined at $n=1$ indicates a
further zero-crossing of the coefficient $k$ at the particle-hole
symmetric point of the Hamiltonian itself. Bottom: cross-sections at
constant interaction $U/\Gamma=5$ and constant occupancy $n=0.6$. }
\label{fig1}
\end{figure}

We therefore solved the impurity problem numerically using the
numerical renormalization group (NRG)
\cite{wilson1975,krishna1980a,krishna1980b,costi1991,hofstetter2000,bulla2008}.
This non-perturbative approach is based on logarithmic discretization
of the continuum, mapping onto a tight-binding chain with
exponentially decreasing hopping constants, and iterative
diagonalization of the resulting Hamiltonian
\cite{wilson1975,krishna1980a,bulla2008}. Through various refinements
over the years
\cite{frota1986,sakai1989,costi1994,bulla1998,hofstetter2000,campo2005,peters2006,weichselbaum2007,resolution}
the technique has developed into a powerful tool for computing
dynamical properties of impurity models. Comparisons with quantum
Monte Carlo simulations indicate that the results of the NRG, when
taken to full convergence, may be considered as essentially exact (up
to very small systematic errors due to discretization and truncation).
We performed the NRG calculations with a narrow broadening kernel by
averaging over $N_z=32$ interleaved discretization grids with the
discretization parameter $\Lambda=2$, and increasing the truncation
cutoff until convergence \cite{resolution}: these steps reduced the
oscillatory artifacts and allowed a reliable extraction of the cubic
term in the self-energy function in the limit of small $\omega$. The
results, shown in Fig.~\ref{fig1}, reveal a change of sign of the
coefficient $b$ in $\Im\Sigma$ along a curve in the $U$-$n$ plane
(black line in the figure). At low $U \ll \Gamma$, the sign reversal
occurs close to half-filling. For $U$ of order $\Gamma$, the
sign-change point rapidly moves away from half-filling. At still
higher $U$, the slope of the black curve in the $U$-$n$ plane
redresses and becomes increasingly vertical for $U \gg \pi \Gamma$.

These results are fully consistent with our perturbative analysis. For
very small $U$, the third order term is negligibly small and the sign
is constant in essentially the full $0<n<1$ interval. In this regime,
the curve separating the different signs in the $U$-$n$ plane is
almost vertical and close to $n=1$. At some value of $U$ of order
$\pi\Gamma$, the perturbation theory predicts that the third-order
contribution will overtake the second-order contribution for most
$E_d$ at almost the same value of $u$, and $b$ will thus change in a
wide $n$ interval. Indeed, this seems to correspond to $U \approx
2\Gamma$ where the curve in the $U$-$n$ plane abruptly changes slope
and becomes almost horizontal. Around the same $u$, however, we enter
the strong-coupling regime where the perturbation theory breaks down.
Note that for large $u=U/\Gamma$ the crossover (as $n$ increases from
0) to the ``strong coupling'' domain with $b>0$ already takes place in
the mixed valent regime of the impurity model; the deep Kondo limit is
confined to values of $n$ close to 1.

\begin{figure}[htbp]
\centering
\includegraphics[width=\columnwidth]{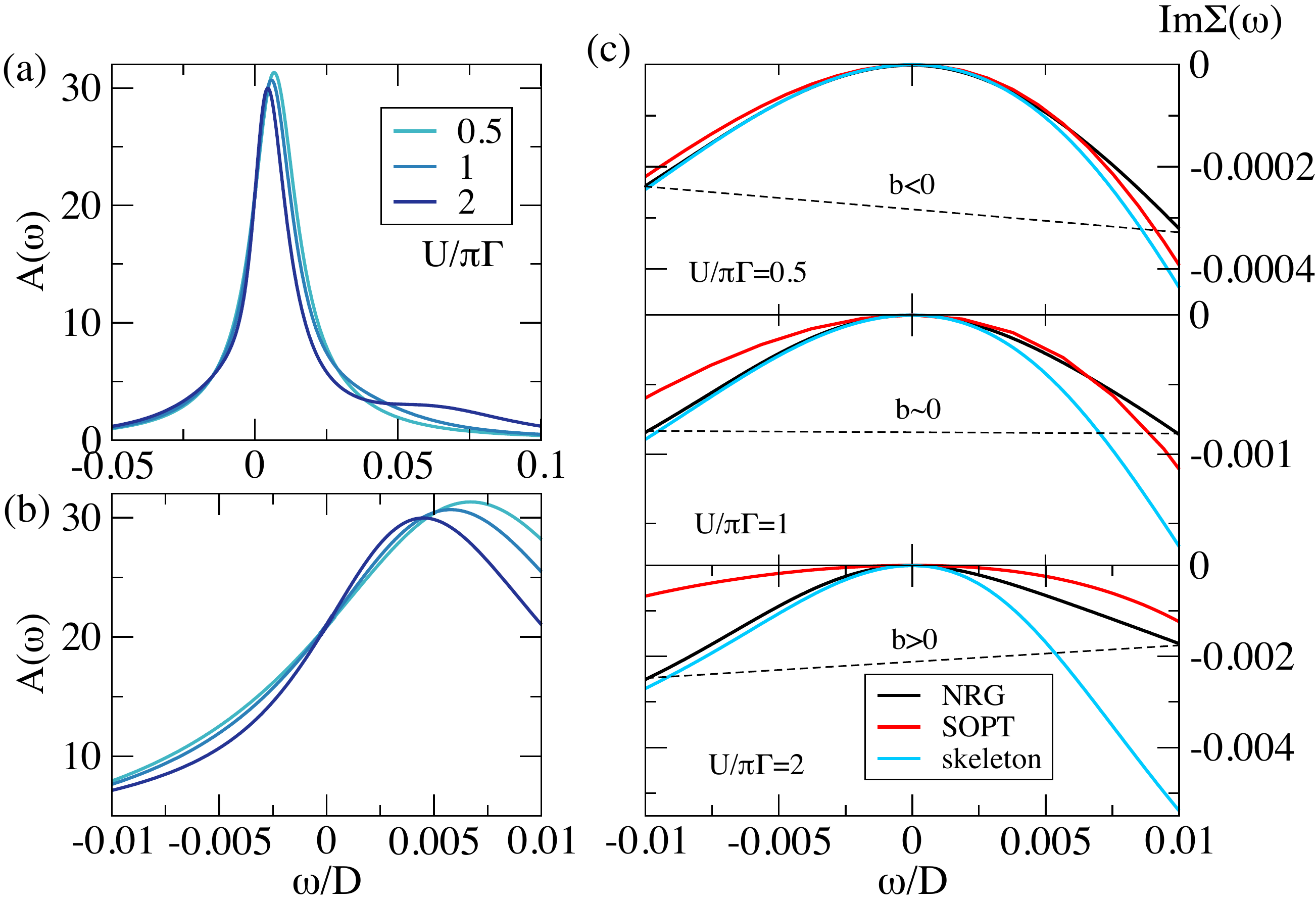}
\caption{Spectral function and imaginary part of self-energy for
$n=0.6$ and three values of $U/\pi\Gamma$ across the sign-reversal
line (magenta line in Fig.~\ref{fig1}). (a) Spectral functions
computed using the NRG, and (b) close-ups on the low-$\omega$ range.
(c) $\Im\Sigma$ computed using the NRG and two weak-coupling
approaches. Here $\Gamma=0.01D$, $T=0$. $b$ for the NRG calculations
has the sign of the slopes of the dashed lines in the figures.}
\label{diff}
\end{figure}

We also performed the skeleton expansion to second order, which is a
self-consistent calculation where the dressed Green's function is used
as the propagator in the second-order term of the self-energy. This
corresponds to an infinite resumation of a certain class of diagrams,
which for small $U$ reduces to the bare perturbation theory in $U$.
The results show that the coefficient $b$ has the same sign for any
value of $U$ and $n$, i.e., the sign associated with the weak-coupling
limit, see Fig.~\ref{diff}. This can also be shown analytically by
invoking the Friedel sum rule \cite{hewson} --- for the skeleton
expansion this leads to $b/a \propto \sin(n\pi)/\Gamma$ which makes $b$
negative for all $n$. The absence of sign change seems to imply that
the skeleton expansion is not able to describe the transition to the
dynamics expected in the strong-coupling regime, presumably because
its starting point is still the non-interacting limit. On the other
hand, the extremely correlated Fermi liquid theory produces the
correct sign of the scattering-rate asymmetry because it is
constructed as a strong-coupling approach by projecting out the
double-occupancy from the outset.

\begin{figure}[htbp]
\centering
\includegraphics[width=\columnwidth]{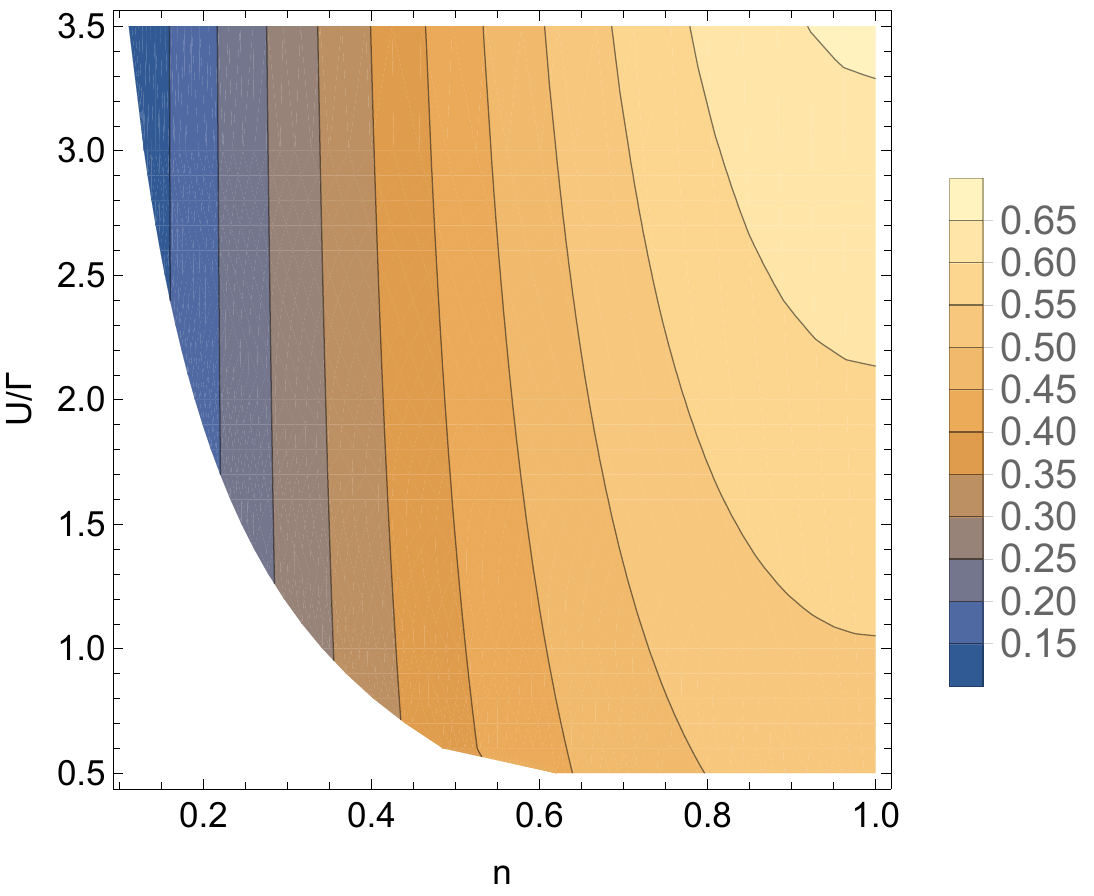}
\caption{Local-moment fraction $f_\mathrm{LM} = \expv{n} -
2\expv{n_\uparrow n_\downarrow}$.}
\label{flm}
\end{figure}

Next, we compare the cross-over curve with other quantities sensitive
to the magnetic behavior of the impurity. The local-moment fraction
$f_\mathrm{LM} = \expv{n} -2 \expv{n_\uparrow n_\downarrow} =
2\expv{n}-\expv{n^2}$ is equal to the expectation value of the
projection operator to the singly-occupied impurity state (the state
which carries the spin degree of freedom). It behaves as
$f_\mathrm{LM} = \expv{n}$ in the $U \gg \Gamma$ limit, while the
small-$U$ dependence is shown in Fig.~\ref{flm}. None of the contours
(iso curves) resembles the sign-reversal curve from Fig.~\ref{fig1}.
Most notably, the contours in $f_\mathrm{LM}$ approach the $n=1$ line
with zero slope and do not curve down. Other thermodynamic quantities,
such as the characteristic low-energy scale of the problem defined by
the temperature where the impurity entropy and the effective moment
become small (i.e., the Kondo temperature in the Kondo regime), also
correlate with the dependence of $f_\mathrm{LM}$ on $U$ and $n$ (not
shown). The scattering asymmetry is thus not simply related to the
degree of local-moment formation, but requires a calculation of
dynamical properties.

We now discuss the relevance of these results to the low-temperature
thermopower in correlated systems. Two cases need to be distinguished.
In quantum dots that are directly described by the SIAM, the
thermopower is determined by the asymmetry of the spectral function
around the Fermi level, i.e., its Fermi-level slope
\cite{scheibner2005, costi2010, andergassen2011}. The asymmetry of the
particle-hole scattering is not important as such, it only enters as
one factor that affects the spectral-function slope. The situation is
different for bulk systems described by the Hubbard model that map
within the dynamical mean field theory (DMFT) approximation to a SIAM
with self-consistently defined hybridisation function
\cite{georges1996,shastry2009,tpsh,haule2009,deng,xu2013,arsenault2013}.
There the thermopower is given by the ``leading'' term proportional to
the Fermi-level slope of the transport function that is ``corrected''
by a term proportional to the coefficients of the cubic terms in
$\Im\Sigma$. This is actually an order 1 correction
\cite{deng,xu2013}, which may become dominant close to half-filling.
Due to the DMFT self-consistency, this term has a complex dependence 
on specific details of the problem (shape of the non-interacting DOS,
doping level, strength of the interaction copared to the critical
$U_{c2}$ of the Mott metal-to-insulator transition). We plan to
present a study of these in a separate publication.

In conclusion, we uncovered a new cross-over line in the phase diagram
of the single-impurity Anderson model with flat hybridisation function
which corresponds to a change in the scattering dynamics. On one side
of this line the impurity behaves as a weakly-renormalized resonant
level, on the other side as a magnetic impurity. On the weakly
correlated side, the particle-like excitations scatter more strongly
than the hole-like excitations, while the opposite is the case on the
strongly correlated side. This cross-over might be directly observable
in quantum dot experiments \cite{goldhabergordon1998a,kretinin2011}
provided that the spectral function can be measured in a sufficient
energy window so that the reconstruction of the full Green's function
$G(\omega)$ is possible via the Kramers-Kronig transformation:
assuming that the hybridization function is approximately constant
close to the Fermi level, the particle-hole scattering asymmetry can
be easily extracted from $\Im[G(\omega)^{-1}]$. The predicted almost
exact particle-hole symmetry of the scattering rate along the
cross-over line in the $U$-$n$ plane should make it an interesting
feature to test for in experiments on magnetic adsorbates and quantum
dots.

\begin{acknowledgments}
R\v{Z} acknowledges the support of the Slovenian Research Agency (ARRS)
under P1-0044 and J1-7259.  The
work at UCSC was supported by the US Department of
Energy (DOE), Office of Science, Basic Energy Sciences
(BES), under Award No. DE-FG02-06ER46319.
HRK acknowledges support from the Department of Science and Technology
(DST), India.
\end{acknowledgments}

\bibliography{paper1}

\end{document}